\documentclass[preprintnumbers,article,amsmath,amssymb,floatfix,10pt,prd,twocolumn,
superscriptaddress,nofootinbib]{revtex4-2}
\usepackage{bm}
\usepackage{amsfonts}
\usepackage{latexsym}
\usepackage[latin1]{inputenc}
\usepackage{graphicx}
\usepackage{amsmath}
\usepackage{palatino}
\usepackage{mathpazo}
\usepackage{textcomp}
\usepackage{float}
\usepackage{booktabs}
\usepackage{dcolumn}
\usepackage{ragged2e}
\usepackage{hyperref}
\hypersetup{colorlinks,citecolor=blue}
\hypersetup{colorlinks=true,linkcolor=red,filecolor=magenta,    urlcolor=blue}
\usepackage{amsmath}
\usepackage{xcolor}
\usepackage{orcidlink}
\usepackage{epsfig}
\usepackage{caption}
\usepackage{subcaption}
\usepackage{commath}
\def\jnl@style{\it}
\def\aaref@jnl#1{{\jnl@style#1}}

\def\aaref@jnl#1{{\jnl@style#1}}

\def\aj{\aaref@jnl{AJ}}                   
\def\apj{\aaref@jnl{ApJ}}                 
\def\apjl{\aaref@jnl{ApJ}}                
\def\apjs{\aaref@jnl{ApJS}}               
\def\apss{\aaref@jnl{Ap\&SS}}             
\def\aap{\aaref@jnl{A\&A}}                
\def\aapr{\aaref@jnl{A\&A~Rev.}}          
\def\aaps{\aaref@jnl{A\&AS}}              
\def\mnras{\aaref@jnl{Mon.~Not.~Roy.~Astron.~Soc.}}             
\def\prd{\aaref@jnl{Phys.~Rev.~D}}        
\def\prc{\aaref@jnl{Phys.~Rev.~C}}  
\def\prl{\aaref@jnl{Phys.~Rev.~Lett.}}    
\def\qjras{\aaref@jnl{QJRAS}}             
\def\skytel{\aaref@jnl{S\&T}}             
\def\ssr{\aaref@jnl{Space~Sci.~Rev.}}     
\def\zap{\aaref@jnl{ZAp}}                 
\def\nat{\aaref@jnl{Nature}}              
\def\aplett{\aaref@jnl{Astrophys.~Lett.}} 
\def\apspr{\aaref@jnl{Astrophys.~Space~Phys.~Res.}} 
\def\physrep{\aaref@jnl{Phys.~Rep.}}      
\def\physscr{\aaref@jnl{Phys.~Scr}}       
\def\commat{\aaref@jnl{Comm.~Math.~Phys.}}              
\def\science{\aaref@jnl{Science}}               
\def\cqg{\aaref@jnl{Classical Quant.~Grav.}}            
\def\jpcs{\aaref@jnl{JPCS}}                                     
\def\ijmpd{\aaref@jnl{Int.~J.~Mod.~Phys.~D}}                    
\def\grg{\aaref@jnl{Gen.~Relat.~Gravit.}}               
\def\rpp{\aaref@jnl{Rep.~Prog.~Phys.}}          
\def\npa{\aaref@jnl{Nucl.~Phys.~A}}        
\def\lrr{\aaref@jnl{Living Rev.~Rel.}}                   
\def\jcap{\aaref@jnl{J.~Cosmology Astropart.~Phys.}}    
\def\rmp{\aaref@jnl{Rev.~Mod.~Phys.}}   
\def\epjc{\aaref@jnl{Eur.~Phys.~J.~C}} 
\def\plb{\aaref@jnl{~Phy.~Lett.~B}} 
\def\mpla{\aaref@jnl{Mod.~Phy.~Lett.~A}} 
\def\arxiv{\aaref@jnl{arxiv.org}}

\allowdisplaybreaks[1]

\begin{document}
\color{black}       
%
\title{Modified $f(Q)$ gravity models and their cosmological consequences}

\author{Gaurav N. Gadbail\orcidlink{0000-0003-0684-9702}}
\email{gauravgadbail6@gmail.com}
\affiliation{Department of Mathematics, Birla Institute of Technology and Science-Pilani,\\ Hyderabad Campus, Hyderabad-500078, India.}

\author{P.K. Sahoo\orcidlink{0000-0003-2130-8832}}
\email{pksahoo@hyderabad.bits-pilani.ac.in}
\affiliation{Department of Mathematics, Birla Institute of Technology and Science-Pilani,\\ Hyderabad Campus, Hyderabad-500078, India.}

%

\begin{abstract}
In this work, we consider three different $f(Q)$ models, such as power-law, exponential, and logarithmic, to study which model better mimics $\Lambda$CDM evolution theoretically. Henceforth, we determine solutions to the $f(Q)$ gravity field equations in the isotropic and homogeneous universe. Since all the models contain two model parameters, we reduce the degrees of freedom using the first Friedman equation at the present time. Further, we check the behavior of cosmological parameters using the obtained solution to the field equations and compare it with the $\Lambda$CDM model. As a result, the power-law model shows a good match with $\Lambda$CDM model for $\lambda=-1$ and $\lambda=-2$, while the exponential model behaves well for the range $5\le \beta<11$, and the logarithmic model matches for $3.8<\gamma<4.4$.
\\


\end{abstract}

\maketitle

\date{\today}

\section{Introduction}
The identity of dark energy, which is responsible for the observed accelerated expansion of the universe, is one of the most significant unanswered issues in modern cosmology. Einstein's equations can be used to quite precisely explain the dynamics by adding a straightforward cosmological constant term, but the required magnitude is mysterious and unsupported by the laws of fundamental physics.  
Therefore, it is important to investigate alternative approaches or generalizations of fundamental theories of gravity, which can give simple explanation of dark energy. There are many directions one could follow to construct gravitational modifications. In this paper, we focus on modified theory of gravity.\\
A specific modified theory of gravity that has recently caught the attention of cosmologists is so-called $f(Q)$ symmetric teleparallel gravity \cite{Jimenez/2018,cosmofQ}. Motivated by the formulation of $f(R)$ \cite{Buchdahl,Barrow/1983,Capozziello/2002,Felice/2010} and $f(T)$ gravity \cite{Capozziello/2011,Cai/2016,Farrugia/2016,Paliathanasis/2018}, where the Lagrangian of the gravitational field equations is a function of the Ricci scalar $R$ and torsion scalar $T$ of the underlying geometry \cite{Buchdahl}. This theory is a generalization of the symmetric teleparallel gravity \cite{Nester/1999} constructed by replacing geometric variable $Q$, the Lagrangian of symmetric teleparallel gravity, by a general function $f(Q)$. It utilizes the curvatureless and torsionless connection to define the covariant derivative instead of using the Levi-Civita connection of general relativity (GR). Besides, the foundation of this $f(Q)$ theory is the nonmetricity $Q$, which geometrically explains how a vector's length might vary in a parallel transport. 
However, the field equations of the $f(T)$ and $f(Q)$ theory are second-order rather than fourth-order, which gives it an advantage over the $f(R)$ gravity theory. As a result, the development of this $f(Q)$ theory provides an individual beginning point for various modified gravity.
Recently, the $f(Q)$ gravity has led to fascinating applications and produces intriguing cosmic phenomenology at the background level \cite{Q1,Q2,Q3,Q4,Q5,Q6,Q7,Q8,Q9,Q11,Q12,Q13,Q14}. It has also been successfully confronted with a variety of background and perturbation observational data, including the Supernovae type Ia (SNIa), Baryonic Acoustic Oscillations (BAO), Cosmic Microwave Background (CMB), Redshift Space Distortion (RSD), growth data \cite{ob1,ob2,ob3,ob4,ob5,ob6,ob7,ob8}. This confrontation reveals that $f(Q)$ gravity may challenge the conventional $\Lambda$CDM scenario. Moreover, the Big Bang Nucleosynthesis (BBN) restrictions are easily overcome by $f(Q)$ gravity \cite{Anagnostopoulos/2023}. For the first time, Sokoliuk et al. \cite{simranlss} used N-body simulations of $f(Q)$ gravitation to study large-scale structure creation observables in order to evaluate the applicability of the theory to cosmological context. Some other work \cite{gaurav1,gaurav2,Esposito/2022,gaurav3} has been done in $f(Q)$ gravity through reconstruction strategy.\\
In this work, we give theoretical predictions on cosmic observables derived from a modified $f(Q)$ gravity model without taking any extra assumptions. Observationally, we can find only a particular value of model parameters. However, this study expands our understanding by delineating a range of model parameters, allowing for an assessment of their suitability for future investigations into our models. We study the general form of the $f(Q)$ model as $f(Q)=Q+F(Q)$, which is equivalent to GR when $F(Q)=0$. To differentiate it from GR, we choose three $F(Q)$ ansatz models, such as power-law, exponential, and logarithmic. Using these three different models, we determine solutions to the field equations in $f(Q)$ gravity in the cosmological scheme of an isotropic and homogeneous universe. Since all the models contain two model parameters, we reduce the degrees of freedom using the first Friedman equation at the present time. Further, we check the behavior of cosmological parameters using the obtained solution to the field equation and compare it with the $\Lambda$CDM model to check its compatibility with $\Lambda$CDM evolution. Also, we compare these models and demonstrate which model is suitable for present-day and late-time studies from a theoretical point of view.
The outline of this work is as follows: In section \ref{section 2}, we present the mathematical formulation of $f(Q)$ gravity theory. In section \ref{section 3}, we consider three different $f(Q)$ ansatz models and find the corresponding solution. In section \ref{section 4}, we summarized the obtained results.

\section{Formulation of $f(Q)$ gravity theory}
\label{section 2}
In contrast to GR, the theory of symmetric teleparallelism demands a different foundation. Since the Levi-Civita connection, which is torsion-free and nonmetricity-free, is used to obtain curvature in GR, symmetric teleparallel gravity requires a different connection that is both curvature-free and torsion-free. The metric affine connection is the most general connection and it is given by 
\begin{equation}
\hat{\Gamma}^{\,\sigma}_{\,\,\,\mu\nu}=\Gamma^{\,\sigma}_{\,\,\,\mu\nu}+K^{\,\sigma}_{\,\,\,\mu\nu}+L^{\,\sigma}_{\,\,\,\mu\nu}
\end{equation}
where the Levi-Civita connection, contorsion tensor, and disformation tensor are given by
\begin{eqnarray*}
\Gamma^{\,\sigma}_{\,\,\,\mu\nu} &=&\frac{1}{2}g^{\sigma\lambda}\left(\partial_{\mu}g_{\lambda\nu}+\partial_{\nu}g_{\lambda\mu}-\partial_{\lambda}g_{\mu\nu}\right),\\
K^{\,\sigma}_{\,\,\,\mu\nu} &=& \frac{1}{2}T^{\,\sigma}_{\,\,\,\mu\nu}+T^{\,\,\,\,\,\,\,\sigma}_{(\mu\,\,\,\,\,\,\nu)},\\
L^{\,\sigma}_{\,\,\,\mu\nu} &=& -\frac{1}{2}g^{\sigma\lambda}\left(Q_{\mu\lambda\nu}+Q_{\nu\lambda\mu}-Q_{\lambda\mu\nu}\right).
\end{eqnarray*}
The non-metricity tensor $Q_{\sigma\mu\nu}$ is defined as
\begin{equation}
\label{1}
Q_{\sigma\mu\nu}=\nabla_{\sigma}g_{\mu\nu},
\end{equation}
and the corresponding traces are 
\begin{equation}
\label{2}
Q_{\sigma}=Q_{\sigma\,\,\,\,\mu}^{\,\,\,\,\mu}\, ,\,\,\,\,\,\,\,\,\tilde{Q}_{\sigma}=Q^{\mu}_{\,\,\,\,\sigma\mu}\,.
\end{equation}
Aside from that, the superpotential tensor $P_{\,\,\mu\nu}^{\sigma}$ can be written as 
\begin{equation}
\label{3}
4P_{\,\,\mu\nu}^{\sigma}=-Q^{\sigma}_{\,\,\,\,\mu\nu}+2Q^{\,\,\,\,\,\,\sigma}_{(\mu\,\,\,\,\nu)}-Q^{\sigma}g_{\mu\nu}-\tilde{Q}^{\sigma}g_{\mu\nu}-\delta^{\sigma}_{(\mu}\, Q\,_{\nu)},
\end{equation} 
obtaining a non-metricity tensor's trace as 
\begin{equation}
\label{4}
Q=-Q_{\sigma\mu\nu}P^{\sigma\mu\nu}.
\end{equation}
In this work, we study the extension of symmetric teleparallel theory called $f(Q)$ gravity theory, and its considered action is given as  \cite{Jimenez/2018}
\begin{equation}
\label{5}
S=\int \left[\frac{1}{2}f(Q)+\lambda_{\alpha}^{\,\,\,\beta\mu\nu} R^{\alpha}_{\,\,\,\beta\mu\nu}+\lambda_{\alpha}^{\,\,\,\mu\nu} T^{\alpha}_{\,\,\,\mu\nu}+\mathcal{L}_m\right]\sqrt{-g}\,d^4x,
\end{equation}
where $\lambda_{\alpha}^{\,\,\,\beta\mu\nu}$ are the Lagrange multipliers, $f(Q)$ is a freely chosen function of a non-metricity tensor's trace, and $g$ is determinant of $g_{\alpha\beta}$, and $\mathcal{L}_m$ is the matter Lagrangian density.\\
As action \eqref{5} varies with respect to the metric, the gravitational field equation for $f(Q)$ is obtained, and it is written as
\begin{multline}
\label{7}
\frac{2}{\sqrt{-g}}\nabla_{\sigma}\left(f_{Q}\sqrt{-g}\,P^{\sigma}_{\,\,\mu\nu}\right)+\frac{1}{2}f\,g_{\mu\nu}\\
+f_{Q}\left(P_{\mu\sigma\lambda}Q_{\nu}^{\,\,\,\sigma\lambda}-2Q_{\sigma\lambda\mu}P^{\sigma\lambda}_{\,\,\,\,\,\,\nu}\right)=- T_{\mu\nu},
\end{multline}
where $f_Q=\frac{d f}{d Q}$. The energy-momentum tensor for matter is now defined as
\begin{equation}
\label{6}
T_{\mu\nu}\equiv-\frac{2}{\sqrt{-g}}\frac{\delta(\sqrt{-g})\mathcal{L}_m} {\delta g^{\mu\nu}}.
\end{equation}
Also, by varying action \eqref{5} with respect to the connection results in:
\begin{equation}
\label{conn}
    \nabla_{\sigma}\lambda_{k}^{\,\,\,\mu\nu\sigma}+\lambda_{k}^{\,\,\,\mu\nu}=\sqrt{-g}f_Q\,P_{\,k}^{\,\,\,\mu\nu}+H_{\,k}^{\,\,\,\mu\nu},
\end{equation}
where $H_{\,k}^{\,\,\,\mu\nu}=-\frac{1}{2}\frac{\delta\mathcal{L}_m}{\delta\Gamma^{k}_{\,\,\,\mu\nu}}$ is the hypermomentum tensor density.\\
It is possible to simplify Eq. \eqref{conn} by taking into account the antisymmetry property of $\mu$ and $\nu$ in the Lagrangian multiplier coefficients
\begin{equation}
    \nabla_{\mu}\nabla_{\nu} \left(f_{Q}\sqrt{-g}\,P_{\,k}^{\,\,\,\mu\nu}+H_{\,k}^{\,\,\,\mu\nu}\right)=0.
\end{equation}
More specifically, the connection can be parameterized with a collection of functions $\xi^{\alpha}$ as
$\hat{\Gamma}^{\,\sigma}_{\,\,\,\alpha\beta}=\frac{\partial x^{\sigma}}{\partial\xi^{\mu}}\partial_{\alpha}\partial_{\beta}\xi^{\mu}$. The connection equation of motion can be easily calculated by noticing that the variation of the connection with respect to $\xi^{\sigma}$ is equivalent to performing a diffeomorphism so that $\delta_{\xi}\hat{\Gamma}^{\,\sigma}_{\,\,\,\mu\nu}=-\mathcal{L}_{\xi}\hat{\Gamma}^{\,\sigma}_{\,\,\,\mu\nu}=-\nabla_{\mu}\nabla_{\nu}\xi^{\sigma}$, where we have used that the connection is flat and torsion-free \cite{cosmofQ}. Furthermore, in the absence of hypermomentum\footnote{If there is no hypermomentum, this is trivially true. Second, we can assume that the hypermomentum is antisymmetric, in which case our assertion is identically true ($H_{\,k}^{\,\,\,\mu\nu}= 0$). Finally, if $H_{\,k}^{\,\,\,\mu\nu}\neq 0$, we consider the assertion to be the hypermomentum conservation law.} \cite{Beltran/2018}, the connection field equations read as
\begin{equation}
\label{8}
\nabla_{\mu}\nabla_{\nu} \left(f_{Q}\sqrt{-g}\,P_{\,k}^{\,\,\,\mu\nu}\right)=0.
\end{equation}

A geometric explanation of gravity that is entirely equivalent to general relativity is symmetric teleparallel gravity. This equivalence is easily demonstrated in the so-called coincident gauge, for which $\hat{\Gamma}^{\,\sigma}_{\,\,\,\mu\nu}=0$. By applying the constraint that the connection is symmetric, the torsion tensor disappears, and the Levi-Civita connection may be written in terms of the disformation tensor as $\hat{\Gamma}^{\,\sigma}_{\,\,\,\mu\nu}=-L^{\,\sigma}_{\,\,\,\mu\nu}$. As a result, the non-metricity is reduced to $Q_{\sigma\mu\nu}=\partial_{\sigma}g_{\mu\nu}$.

The energy-momentum tensor $T_{\mu\nu}$ of the matter, which is assumed to be a perfect fluid, is given by
\begin{equation}
\label{9}
T_{\mu\nu}=(p+\rho)u_{\mu}u_{\nu}+pg_{\mu\nu},
\end{equation}
where $u_{\mu}$ is a four-velocity vector, $p$ and $\rho$ are the pressure and energy density of a perfect fluid, respectively.\\
We assume a Friedmann-Lemaitre-Robertson-Walker (FLRW) spacetime with a metric of the type that is homogeneous, isotropic, and spatially flat at the background level
\begin{equation}
\label{10}
ds^2=-dt^2+a^2(t)(dx^2+dy^2+dz^2),
\end{equation}
where $a(t)$ is a scale factor. For the FLRW metric, the nonmetricity scalar reads $Q=6H^2$, where $H=\frac{\dot{a}}{a}$ is the Hubble parameter. Applying the FLRW metric and imposing the splitting function $f(Q) = Q + F(Q)$, the relevant Friedman equations are
 
\begin{equation}
\label{f1}
3 H^2=\rho +\frac{F}{2}-Q\,F_Q,
\end{equation} 
\begin{equation}
\label{f2}
\left(2Q\,F_{QQ}+F_Q+1\right)\dot{H}+\frac{1}{4}\left(Q+2Q\,F_Q-F\right)=-2p,
\end{equation}
where $F_Q=\frac{dF}{dQ}$, and $F_{QQ}=\frac{d^2F}{dQ^2}$, dot $(.)$ denotes derivative with respect to $t$. \\
In this scenario, we investigate the possibility that the Universe is filled with dust and radiation fluids, and hence
\begin{equation}
    \rho=\rho_m+\rho_r,\,\,\,\,\,p=p_m+p_r=\frac{1}{3}\rho_r,
\end{equation}
where $\rho_r$ and $\rho_m$ are the energy densities of the radiation and dust, respectively. Then Eqs.\eqref{f1} and \eqref{f2} can be written as
\begin{equation}
    3H^2=\rho_m+\rho_r+\rho_Q,
\end{equation}
\begin{equation}
    2\dot{H}+3H^2=-\frac{\rho_r}{3}-p_Q,
\end{equation}
where $p_Q$ and $\rho_Q$ are the pressure and DE density, respectively, which is contributed to the geometry given by
\begin{equation}
    \rho_Q=\frac{F}{2}-Q\,F_Q,
\end{equation}
\begin{equation}
    p_Q=2\dot{H}\left(2Q\,F_{QQ}+F_Q\right)-\rho_Q.
\end{equation}
As a result, the equation of state due to DE is given by 
\begin{equation}
\label{w}
    \omega_Q=\frac{p_Q}{\rho_Q}=-1+\frac{4\dot{H}\left(2Q\,F_{QQ}+F_Q\right)}{F-2Q\,F_Q}.
\end{equation}
Further, we would like to mention here that the $f(Q)$ gravity satisfies the conservation equation
\begin{equation}
\dot{\rho}+3H(\rho+p)=0.
\label{c}
\end{equation}

\section{Three different $f(Q)$ ansatz models}
\label{section 3}
\subsection{Power-law $f(Q)$ model}
\begin{widetext}

\begin{figure}
\centering
\subcaptionbox{}
{\includegraphics[width=0.31\textwidth]{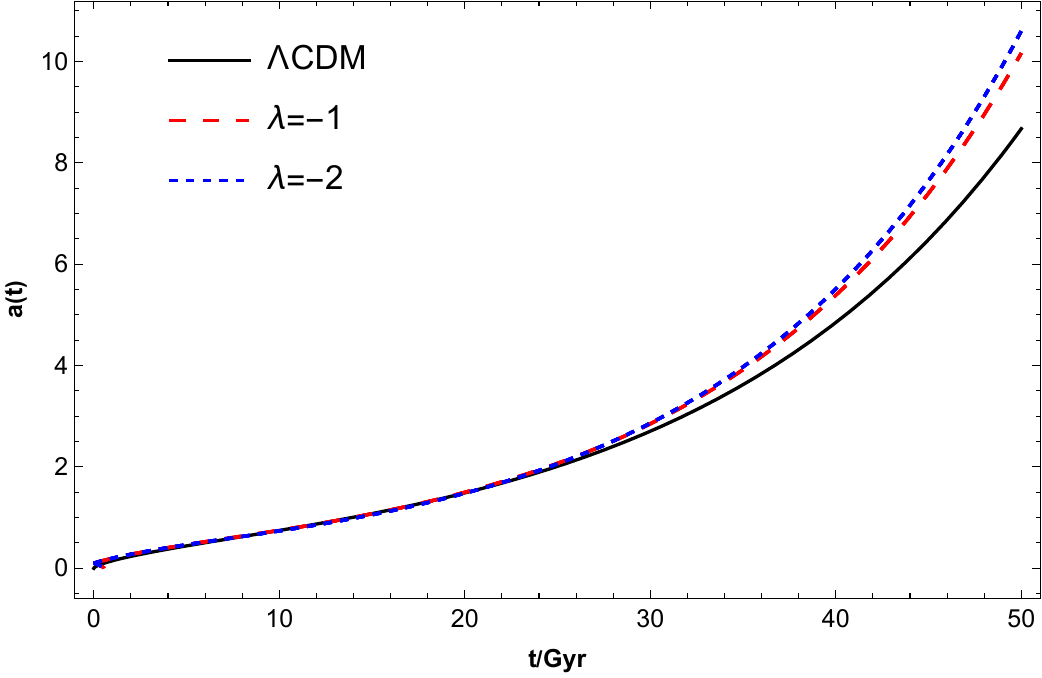}}
 \hspace{0.15in} 
\subcaptionbox{}{\includegraphics[width=0.31\textwidth]{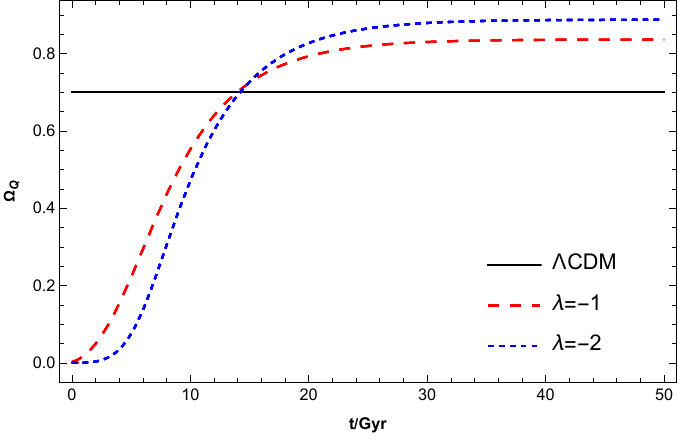}}%
 \hspace{0.15in} 
\subcaptionbox{}{\includegraphics[width=0.31\textwidth]{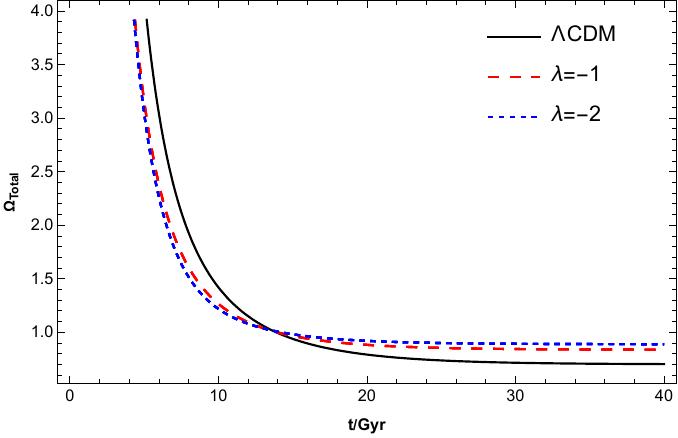}}%
\hspace{0.15in} 
\subcaptionbox{}{\includegraphics[width=0.31\textwidth]{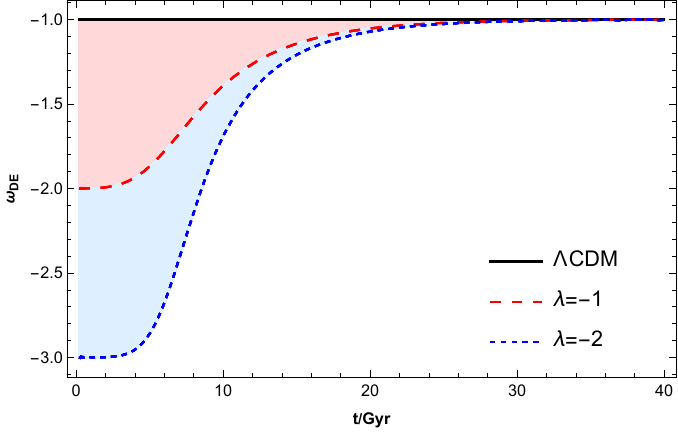}}%
\hspace{0.15in} 
\subcaptionbox{}{\includegraphics[width=0.31\textwidth]{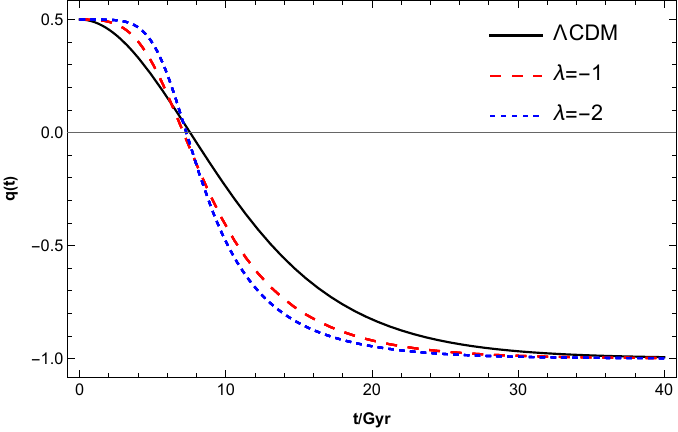}}%
\hspace{0.15in} 
\subcaptionbox{}{\includegraphics[width=0.31\textwidth]{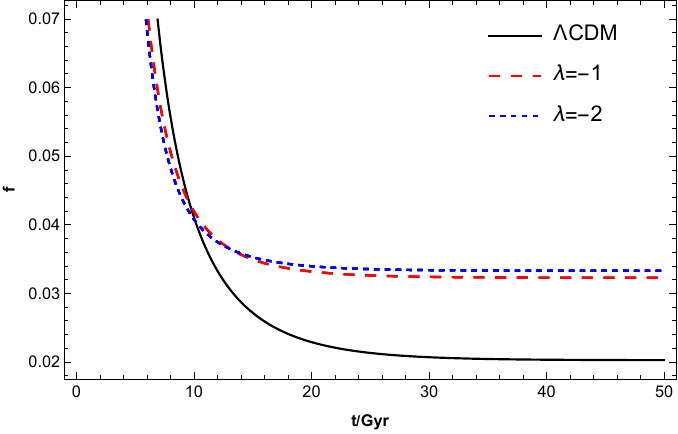}}%
\\ 
\caption{\justifying
Plot (a), (b), (c), (d), and (e) show the evolution of the scale factor $a(t)$, DE density parameter $\Omega_Q$, total density parameter $\Omega_{Total}$, DE EoS parameter $\omega_Q$, and deceleration parameter $q(t)$ with cosmic time $t$, respectively. Plot (f) shows the evolution of the power-law function $f$ with cosmic time $t$.}
\label{M1}
\end{figure}

\end{widetext}

First, we consider the power-law $F(Q)$ model \cite{cosmofQ,Lazkoz/2019,Khyllep/2023}, which is of the form $F(Q)=\alpha \left(\frac{Q}{Q_0}\right)^{\lambda}$, where $\alpha$ and $\lambda$ are model parameters. For $\alpha=0$ the model reduces to the symmetric teleparallel theory equivalent to GR. When $\lambda=0$, the model reduces to $f(Q)=Q+2\Lambda(=\alpha)$, which recover the $\Lambda$CDM expansion history of the universe.
However, for $\lambda<1$ and $\lambda>1$, modification from the GR evolution happens in the low curvatures regime and the high curvatures regime, respectively. Therefore, models with $\lambda<1$ will be applicable to the late-time DE-dominated universe, whereas models with $\lambda>1$ will be appropriate to the early universe. This model is capable of describing late-time universe acceleration and is also compatible with BBN limitations \cite{Anagnostopoulos/2023}. 
For more simplicity, we also considered matter case as $\omega=0$.\\
Using power-law $F(Q)$ model and Eq. \eqref{f1}, we obtained 
\begin{equation}
\label{22}
    \frac{Q}{Q_0}+\frac{\alpha}{Q_0}(2\lambda-1)\left(\frac{Q}{Q_0}\right)^{\lambda}=\frac{\Omega_{m,0}}{a(t)^3},
\end{equation}
where $\Omega_{m,0}$ is the present matter density, $a(t)$ is the cosmic scale factor, and $Q_0=6H_0^2$.\\
Using the above equation, the value of model parameter $\alpha$ at the present time is obtained as 
\begin{equation}
    \alpha=\frac{(\Omega_{m,0}-1)Q_0}{2\lambda-1},\,\,\,\,\,\,\,\,\,\lambda\neq \frac{1}{2}.
\end{equation}
Using the value of $\alpha$, Eq. \eqref{22} further reduce into 
\begin{equation}
\label{24}
    \frac{Q}{Q_0}+(\Omega_{m,0}-1)\left(\frac{Q}{Q_0}\right)^{\lambda}=\frac{\Omega_{m,0}}{a(t)^3}.
\end{equation}
In this case, DE density is expressed as 

\begin{equation}
    2\,\rho_Q=Q_0 (1-\Omega_{m,0})\left(\frac{Q}{Q_0}\right)^{\lambda}.
\end{equation}
The DE density parameter is thus found to be\footnote{The density parameter of dark energy (DE) is determined as follows: We utilize the critical density value as of today to define density parameters. In our calculation, we incorporate the current value of the density parameter and consider its evolution using the $(1+z)$ power term. This approach is crucial as we possess information solely about the current density parameter value, while inferring past values is significantly more challenging \cite{Ghose/2012}.}
\begin{equation}
    \Omega_Q(t)=\frac{\rho_Q}{\rho_{cr}}=(1-\Omega_{m,0})\left(\frac{Q}{Q_0}\right)^{\lambda}.
\end{equation}
Using Eq. \eqref{w}, the DE EoS parameter for the power-law $F(Q)$ model becomes
\begin{equation}
    \omega_Q=-1+\frac{2\lambda}{3}\frac{\dot{H}}{H^2},
\end{equation}
which shows the phantom dominated universe when $\omega_Q<-1$, $\Lambda$CDM universe when $\omega_Q=-1$, and quintessence dominated universe when $-1<\omega_Q<-\frac{1}{3}$.\\
The deceleration parameter for power-law $F(Q)$ model takes the form
\begin{equation}
    q(t)=-1+\frac{3}{2}\left(\frac{Q+\alpha(2\lambda-1)\left(\frac{Q}{Q_0}\right)^{\lambda}}{Q+\alpha\lambda(2\lambda-1)\left(\frac{Q}{Q_0}\right)^{\lambda}}\right),
\end{equation}
which shows the accelerating universe when $q(t)<0$, and decelerating universe when $q(t)>0$.\\
For this power-law model, we try to choose the value of $\lambda$ strictly less than one because it applies to the late-time DE-dominated universe. First, we attempt to determine the value of $\lambda$ between 0 and 1 (except $\lambda\neq \frac{1}{2}$ because the parameter $\alpha$ is not defined at $\lambda=1/2$). For this value of $\lambda$, the scale factor exhibits the expansion behavior, and the deceleration parameter shows the acceleration expansion of the universe, but other cosmological parameters do not demonstrate good compatibility with the $\Lambda$CDM model. For $\lambda=0$, this power-law model exhibits the $\Lambda$CDM expansion history of the universe. As a result, we consider the value $\lambda<0$, and all cosmological parameters produce a good result with good $\Lambda$CDM model compatibility for $\lambda=-1$ and $\lambda=-2$.\\
To obtain the scale factor, we solve Eq. \eqref{24}, assuming $\Omega_{m,0}=0.3$ and $H_0=(14.4)^{-1}$, and we get a numerical solution for $\lambda=-1$ and $\lambda=-2$. In Fig. \ref{M1}(a), we compare the behavior of the scale factor with the $\Lambda$CDM model, and it exhibits good compatibility with the $\Lambda$CDM model. Next, we display the dark energy density parameter $\Omega_Q(t)$ with respect to time $t$ in Fig. \ref{M1}(b), which is not constant. However, the behavior of the DE density parameter is constant in the $\Lambda$CDM model. But, at the present time, the value of $\Omega_Q$ for both models is $0.697$, which is an excellent match with the $\Lambda$CDM value. The DE density parameter $\Omega_Q(t)$ eventually converges to $0.85$ for $\lambda=-1$ and $0.9$ for $\lambda=-2$. Also, for both power-law models, the total density parameter $\Omega_{Total}$ decreases along with the $\Lambda$CDM model (see in fig. \ref{M1}(c)), and it yields the $\Omega_{Total}(t_0)\approx 1$ at the present time.\\

The evolution trajectory of the DE equation-of-state (EoS) parameter $\omega_{DE}$ with respect to $t$ is shown in Fig. \ref{M1}(d), which is not constant. However, the evolution trajectory of the DE EoS parameter is constant in the $\Lambda$CDM model. The whole trajectory of both models is in the phantom regime, and the present value of $\omega_{DE}$ is $-1.18$ for $\lambda=-1$ and $-1.25$ for $\lambda=-2$, and at late-time, it converges to $-1$ ($\Lambda$CDM model). In Fig. \ref{M1}(e), we can see the evolution trajectory of the deceleration parameter $q(t)$ with respect to time $t$. This trajectory $q(t)$ shows the transition of the universe from the deceleration phase to the acceleration phase. At the present time, it shows the acceleration expansion of the universe, and at late time, it will converge to $-1$.\\
Lastly, we plot the power-law model with respect to time $t$ using obtained scale factor for $\Lambda$CDM case, $\lambda=-1$ and $\lambda=-2$ (see in Fig. \ref{M1}(f)). Our two power-law models decay with time $t$, similar to the $\Lambda$CDM model, and we get modest deviations from the $\Lambda$CDM model as time goes on.

\subsection{Exponential $F(Q)$ model}

\begin{widetext}

\begin{figure}
\centering
\subcaptionbox{}
{\includegraphics[width=0.31\textwidth]{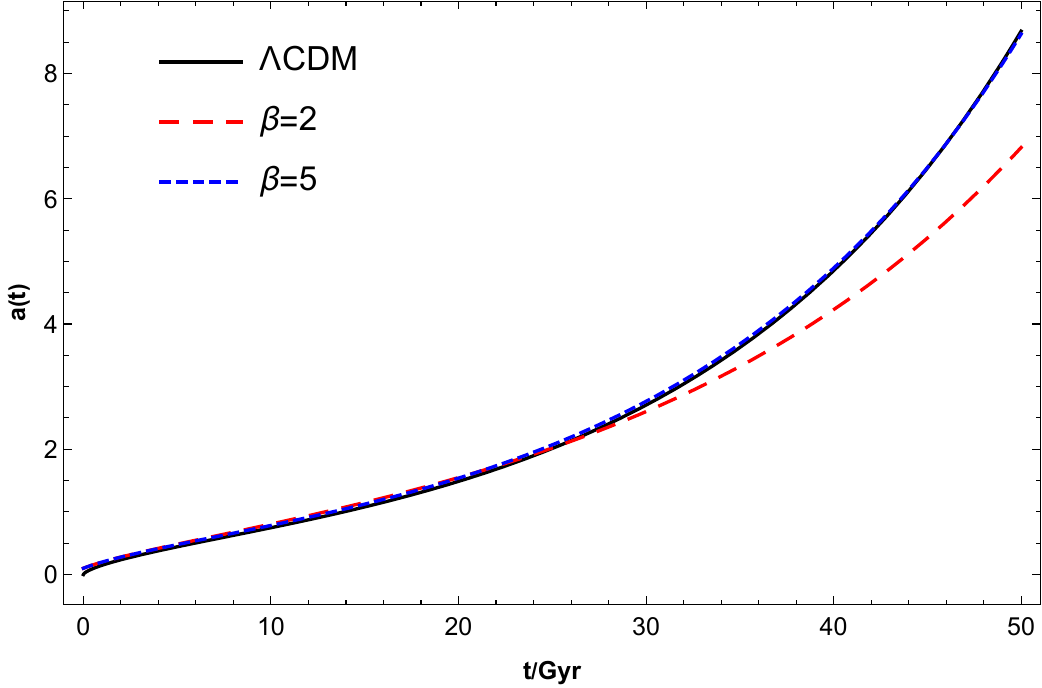}}
 \hspace{0.15in} 
\subcaptionbox{}{\includegraphics[width=0.31\textwidth]{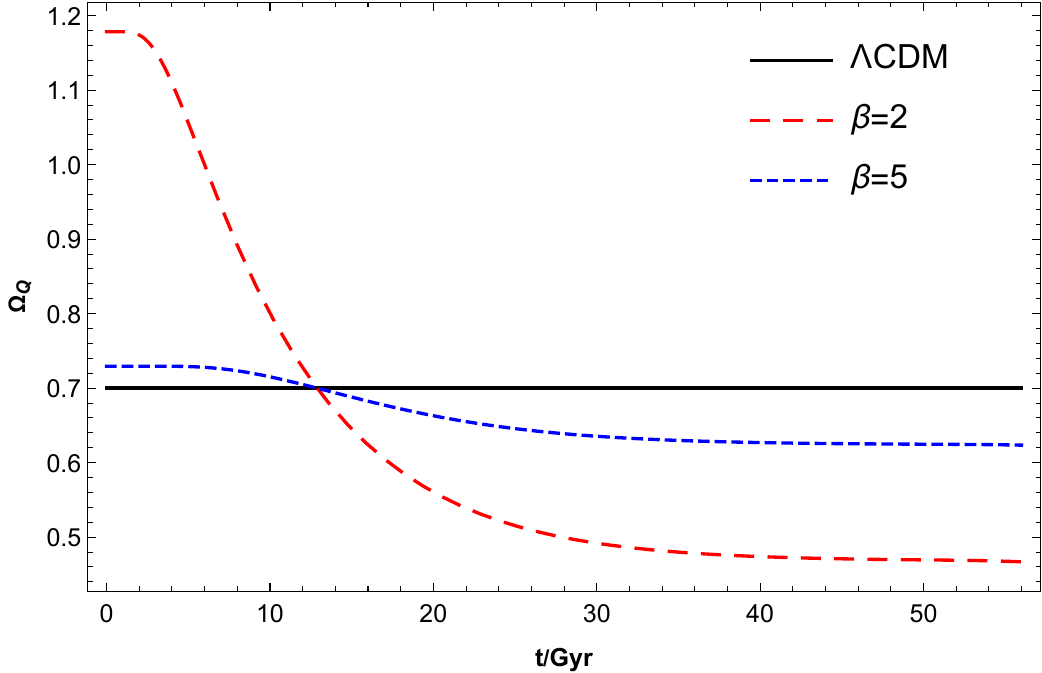}}%
 \hspace{0.15in} 
\subcaptionbox{}{\includegraphics[width=0.31\textwidth]{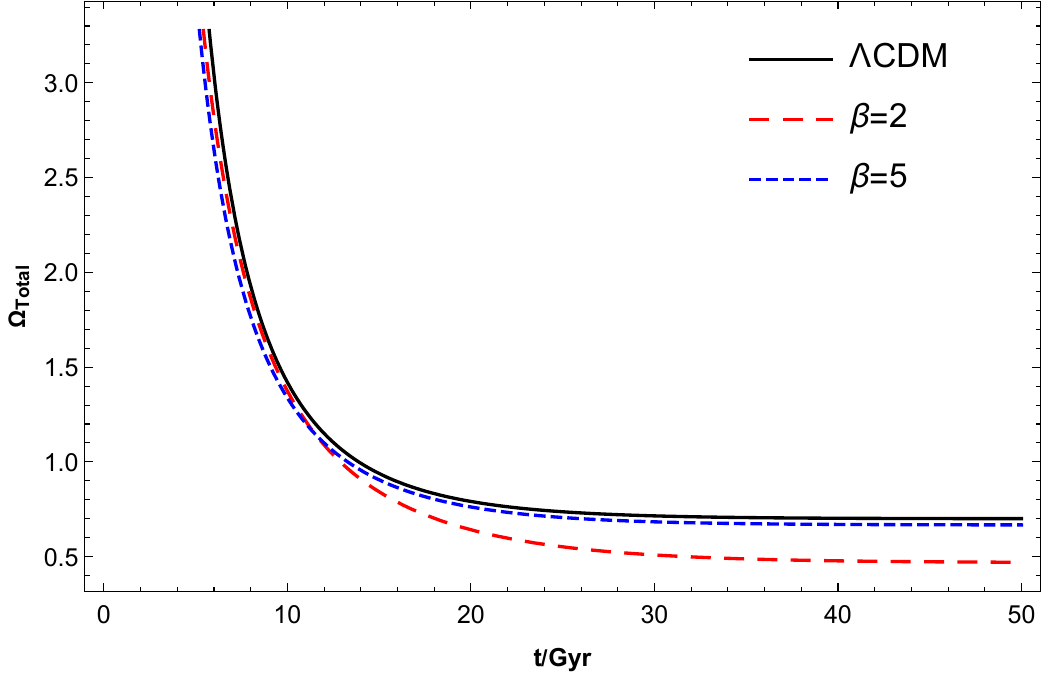}}%
\hspace{0.15in} 
\subcaptionbox{}{\includegraphics[width=0.31\textwidth]{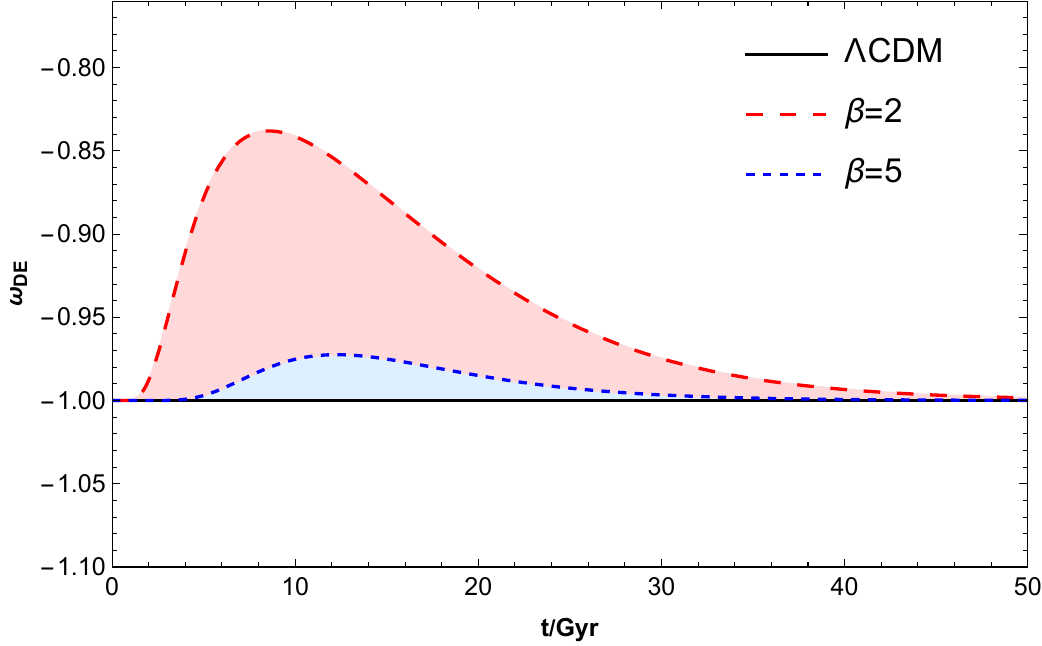}}%
\hspace{0.15in} 
\subcaptionbox{}{\includegraphics[width=0.31\textwidth]{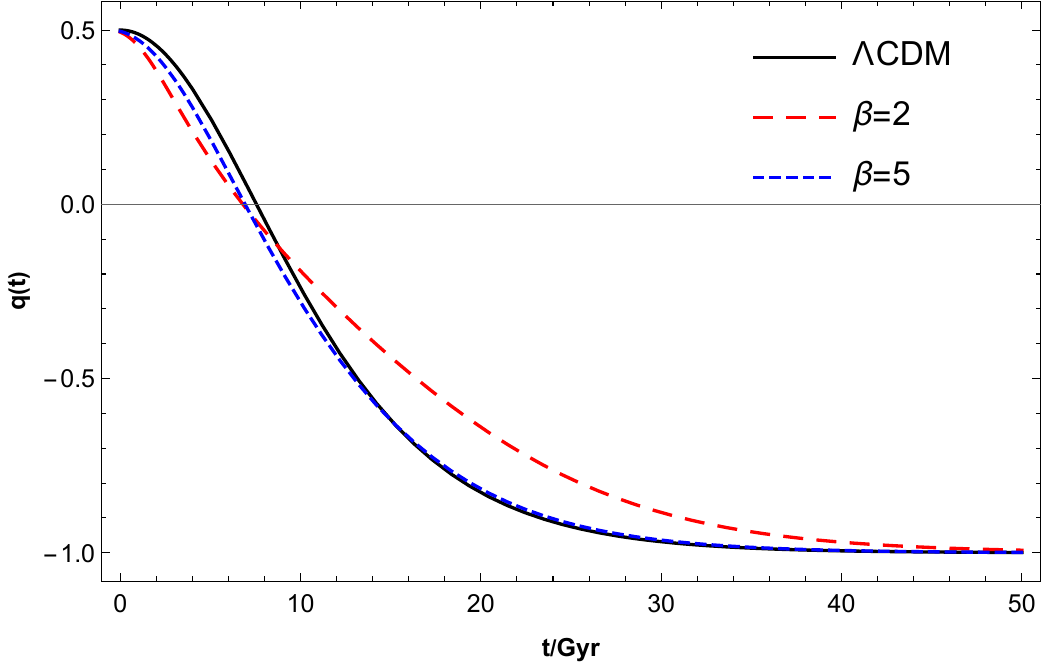}}%
\hspace{0.15in} 
\subcaptionbox{}{\includegraphics[width=0.31\textwidth]{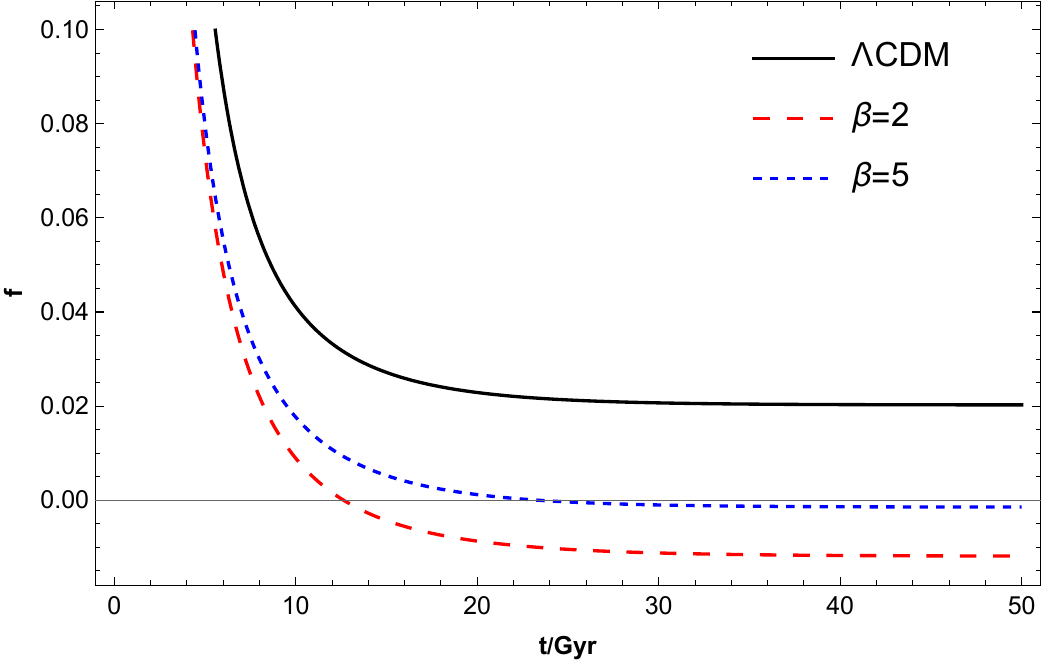}}%
\\ 
\caption{\justifying
Plot (a), (b), (c), (d), and (e) show the evolution of the scale factor $a(t)$, DE density parameter $\Omega_Q$, total density parameter $\Omega_{Total}$, DE EoS parameter $\omega_Q$, and deceleration parameter $q(t)$ with cosmic time $t$, respectively. Plot (f) shows the evolution of the power-law function $f$ with cosmic time $t$. }
\label{M2}
\end{figure}
\end{widetext}

Next, we consider the exponential $F(Q)$ model \cite{simranlss}, which is of the form $F(Q)=\alpha Q_0\left(1-exp\left(-\beta\sqrt{\frac{Q}{Q_0}}\right)\right)$, where $\alpha$ and $\beta$ are model parameters. For $\beta=0$ the model reduces to the symmetric teleparallel theory equivalent to GR without a cosmological constant. 
Numerous cosmic possibilities have been investigated in $f(Q)$ theory utilizing various exponential models, most notably inflationary cosmology, BBN constraints, and dynamic system analysis \cite{Anagnostopoulos/2023,Khyllep/2023}.
For more simplicity, we also considered matter case as $\omega=0$.\\
Using exponential $F(Q)$ model and Eq. \eqref{f1}, we obtained 
\begin{equation}
\label{29}
    \frac{Q}{Q_0}-\alpha\left[1-\left(1+\beta\sqrt{\frac{Q}{Q_0}}\right)exp\left(-\beta\sqrt{\frac{Q}{Q_0}}\right)\right]=\frac{\Omega_{m,0}}{a(t)^3},
\end{equation}
where $\Omega_{m,0}$ is the present matter density, $a(t)$ is the cosmic scale factor, and $Q_0=6H_0^2$.\\
Using the above equation, the value of model parameter $\alpha$ at the present time is obtained as 
\begin{equation}
    \alpha=\frac{1-\Omega_{m,0}}{1-(1+\beta)\,exp(-\beta)}.
\end{equation}
Using the value of $\alpha$, Eq. \eqref{29} further reduce into 
\begin{multline}
\label{31}
    \frac{Q}{Q_0}-\frac{1-\Omega_{m,0}}{1-(1+\beta)\,exp(-\beta)}\\
    \left[1-\left(1+\beta\sqrt{\frac{Q}{Q_0}}\right)exp\left(-\beta\sqrt{\frac{Q}{Q_0}}\right)\right]=\frac{\Omega_{m,0}}{a(t)^3}.
\end{multline}
In this case, DE density is expressed as 

\begin{multline}
   2\,\rho_Q=\frac{(1-\Omega_{m,0})Q_0}{1-(1+\beta)\,exp(-\beta)}\\
    \left[1-\left(1+\beta\sqrt{\frac{Q}{Q_0}}\right)exp\left(-\beta\sqrt{\frac{Q}{Q_0}}\right)\right].
\end{multline}
The DE density parameter is thus found to be
\begin{multline}
   \Omega_Q(t)=\frac{1-\Omega_{m,0}}{1-(1+\beta)\,exp(-\beta)}\\
    \left[1-\left(1+\beta\sqrt{\frac{Q}{Q_0}}\right)exp\left(-\beta\sqrt{\frac{Q}{Q_0}}\right)\right].
\end{multline}
Using Eq. \eqref{w}, the DE EoS parameter for the exponential $F(Q)$ model becomes
\begin{multline}
    \omega_Q=-1+\frac{2\dot{H}}{Q_0}\\
    \left(\frac{\beta^2}{exp(\beta\sqrt{\frac{Q}{Q_0}})-1-\frac{1-\Omega_{m,0}}{1-(1+\beta)\,exp(-\beta)}\sqrt{Q\,Q_0}} \right).
\end{multline}
The deceleration parameter for the exponential $F(Q)$ model takes the form
\begin{multline}
    q(t)=-1+\frac{3}{2}\times\\
    \frac{1-\frac{1-\Omega_{m,0}}{1-(1+\beta)\,exp(-\beta)}\frac{Q_0}{Q}\left[1-\left(1+\beta\sqrt{\frac{Q}{Q_0}}\right)exp\left(-\beta\sqrt{\frac{Q}{Q_0}}\right)\right]}{1-\frac{1-\Omega_{m,0}}{1-(1+\beta)\,exp(-\beta)}\frac{\beta^2}{2}\,exp(-\beta\sqrt{\frac{Q}{Q_0}})}.
\end{multline}
For the exponential model, the numerical solution of the scale factor is obtained only for $0<\beta<11$, which depicts the expansion behavior of the universe. As a result, we try to choose the value of $\beta$ between 0 to 11. When we increase the value of $\beta$ from 0 to 11, our exponential model becomes more compatible with the $\Lambda$CDM model. So we pick two random $\beta$ values: $\beta=2$ (low value) and $\beta=5$ (high value). These two values clearly show the exponential model's change from 0 to 11.\\
To obtain the scale factor, we solve Eq. \eqref{31}, assuming $\Omega_{m,0}=0.3$ and $H_0=(14.4)^{-1}$, and we get a numerical solution for $\beta=2$ and $\beta=5$. In Fig. \ref{M2}(a), we compare the behavior of the scale factor with the $\Lambda$CDM model, and the model for $\beta=5$ exhibits good compatibility with the $\Lambda$CDM model as compared to the model for $\beta=2$. Next, we display the dark energy density parameter $\Omega_Q(t)$ with respect to time $t$ in Fig. \ref{M2}(b), which is not constant. However, the behavior of the DE density parameter is constant in the $\Lambda$CDM model. But, at the present time, the value of $\Omega_Q$ is $0.668$ and $0.692$ for $\beta=2$ and $\beta=5$, respectively, which is a good match with the $\Lambda$CDM value. Model for $\beta=5$ exhibits close behavior to $\Lambda$CDM model. The DE density parameter $\Omega_Q(t)$ eventually converges to $0.46$ for $\beta=2$ and $0.62$ for $\beta=5$ at late time.
Also, for both exponential models, the total density parameter $\Omega_{Total}$ decreases along with the $\Lambda$CDM model (see in fig. \ref{M2}(c)), and it yields the $\Omega_{Total}(t_0)\approx 0.89$ for $\beta=2$ and $\Omega_{Total}(t_0)\approx 0.96$ for $\beta=5$ at the present time.\\
The evolution trajectory of the DE equation-of-state (EoS) parameter $\omega_{DE}$ with respect to $t$ is shown in Fig. \ref{M2}(d), which is not constant. However, the evolution trajectory of the DE EoS parameter is constant in the $\Lambda$CDM model. The evolution trajectory of both models starts with the $\Lambda$CDM model, exhibiting the variation from the $\Lambda$CDM model in the quintessence regime. The present value of $\omega_{DE}$ is $-0.87$ for $\beta=2$ and $-0.97$ for $\beta=5$, and at late-time, it converges to $-1$ ($\Lambda$CDM model). In Fig. \ref{M2}(e), we can see the evolution trajectory of the deceleration parameter $q(t)$ with respect to time $t$. This trajectory $q(t)$ shows the transition of the universe from the deceleration phase to the acceleration phase. At the present time, it shows the acceleration expansion of the universe, and at late time, it will converge to $-1$.\\
Lastly, we plot the exponential model with respect to time $t$ using the obtained scale factor for $\Lambda$CDM case, $\beta=2$, and $\beta=5$ (see in Fig. \ref{M2}(f)). Our two exponential models decay with time $t$, similar to the $\Lambda$CDM model. As time goes on, we observe little more variation from the $\Lambda$CDM model for the $\beta=2$ model compared to the $\beta=5$ model.

\subsection{Logarithmic $F(Q)$ model}
\begin{widetext}

\begin{figure}
\centering
\subcaptionbox{}
{\includegraphics[width=0.31\textwidth]{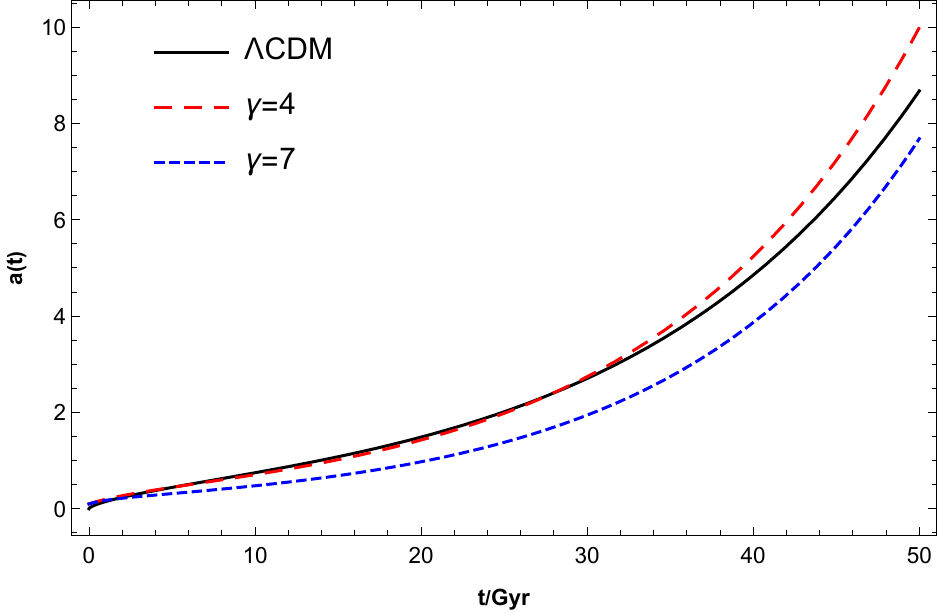}}
 \hspace{0.15in} 
\subcaptionbox{}{\includegraphics[width=0.31\textwidth]{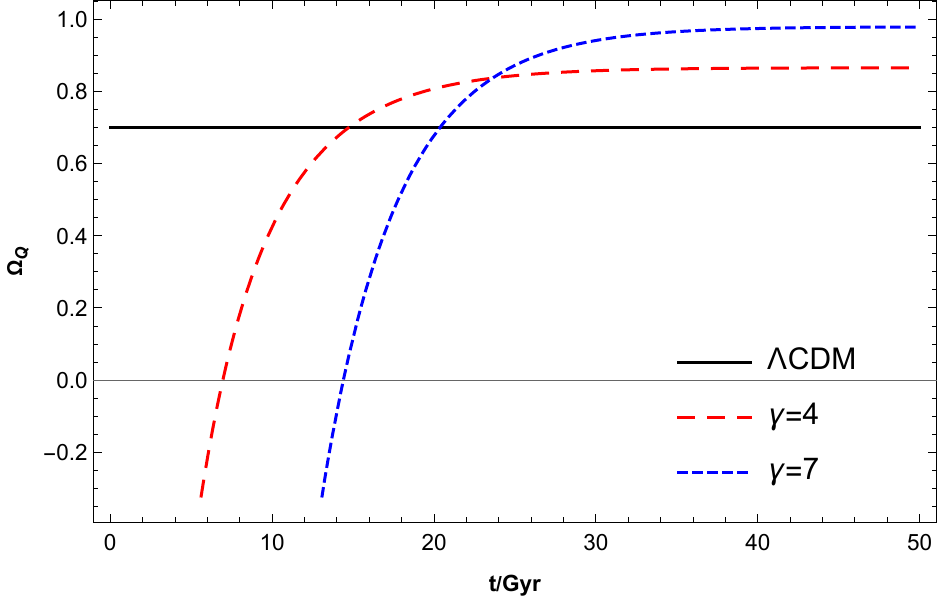}}%
 \hspace{0.15in} 
\subcaptionbox{}{\includegraphics[width=0.31\textwidth]{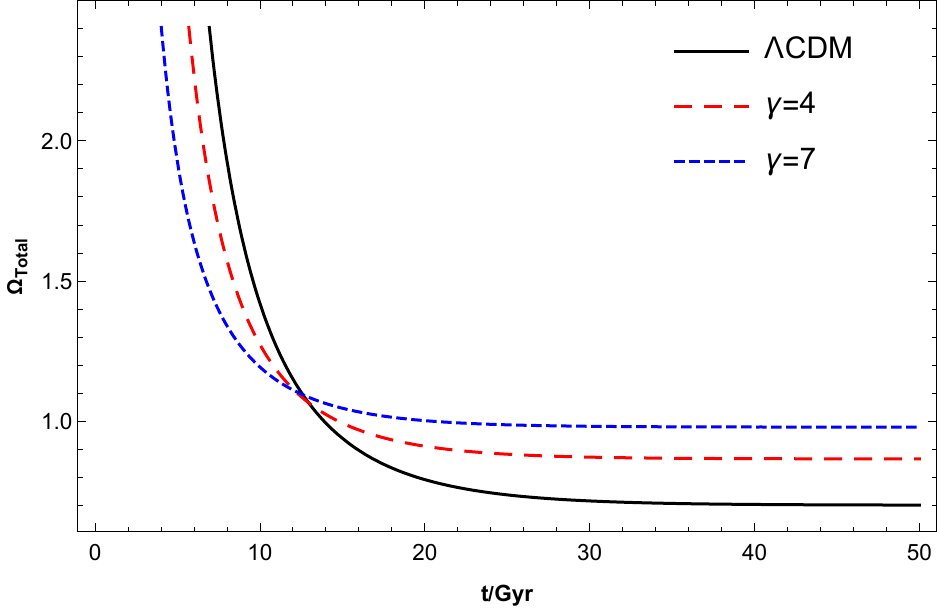}}%
\hspace{0.15in} 
\subcaptionbox{}{\includegraphics[width=0.31\textwidth]{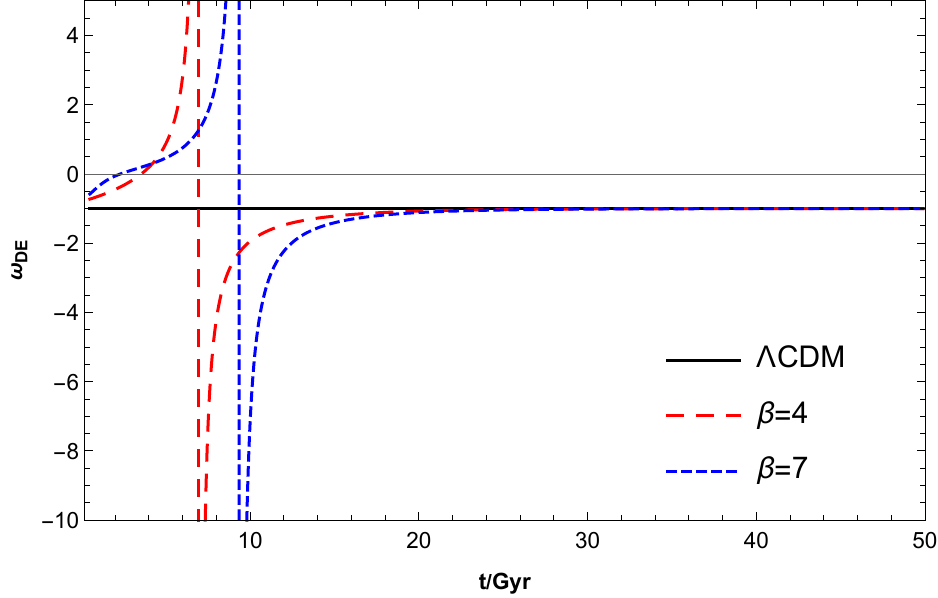}}%
\hspace{0.15in} 
\subcaptionbox{}{\includegraphics[width=0.31\textwidth]{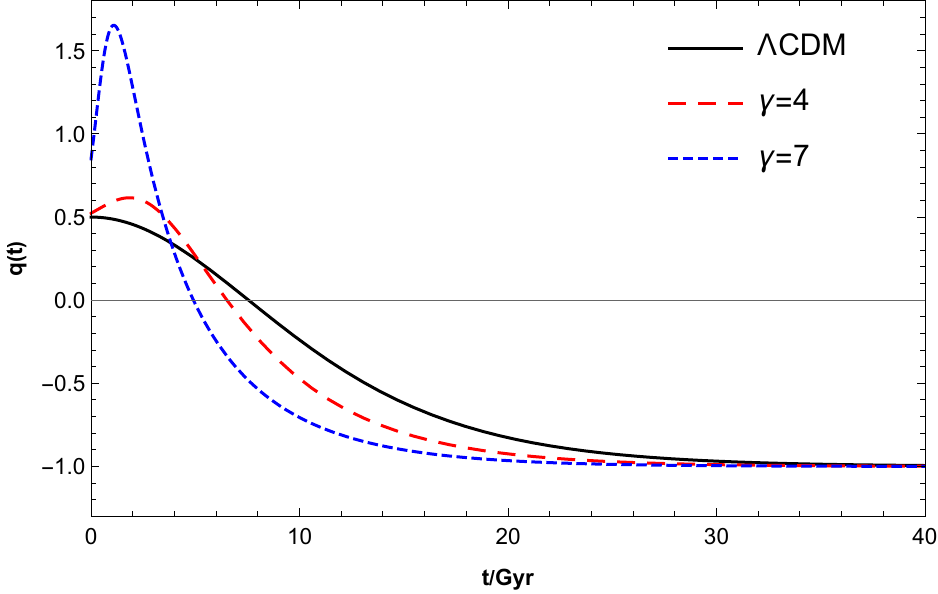}}%
\hspace{0.15in} 
\subcaptionbox{}{\includegraphics[width=0.31\textwidth]{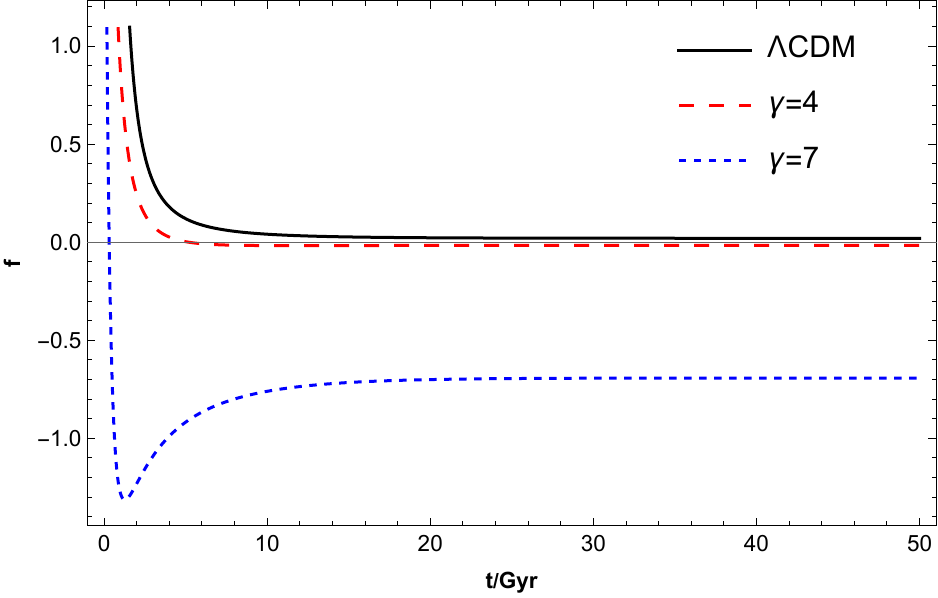}}%
\\ 
\caption{\justifying
Plot (a), (b), (c), (d), and (e) show the evolution of the scale factor $a(t)$, DE density parameter $\Omega_Q$, total density parameter $\Omega_{Total}$, DE EoS parameter $\omega_Q$, and deceleration parameter $q(t)$ with cosmic time $t$, respectively. Plot (f) shows the evolution of the power-law function $f$ with cosmic time $t$.}
\label{M3}
\end{figure}

\end{widetext}

Last, we consider the Logarithmic $F(Q)$ model \cite{Najera/2023}, which is of the form $F(Q)=\alpha\, ln\left(\gamma \frac{Q}{Q_0}\right)$, where $\alpha$ and $\gamma>0$ are model parameters. For $\alpha=0$ the model reduces to the symmetric teleparallel theory equivalent to GR without a cosmological constant. When tested with SNeIa and BAO data, this model effectively predicted the cosmic late-time acceleration in geometrical means \cite{Najera/2023}. In $f(Q)$ theory, a variety of cosmic scenarios have been examined using a variety of logarithmic models \cite{Q11,Anagnostopoulos/2023}. For more simplicity, we also considered matter case as $\omega=0$.\\
Using power-law $f(Q)$ model and \eqref{f1}, we obtained 
\begin{equation}
\label{36}
    \frac{Q}{Q_0}+\frac{\alpha}{Q_0}\left(2-ln\left(\gamma \frac{Q}{Q_0}\right)\right)=\frac{\Omega_{m,0}}{a(t)^3},
\end{equation}
where $\Omega_{m,0}$ is the present matter density, $a(t)$ is the cosmic scale factor, and $Q_0=6H_0^2$.\\
Using the above equation, the value of model parameter $\alpha$ at the present time is 
\begin{equation}
    \alpha=\frac{(\Omega_{m,0}-1)\,Q_0}{2-ln(\gamma)}.
\end{equation}
Using the value of $\alpha$, Eq. \eqref{36} further reduce into 
\begin{equation}
\label{38}
    \frac{Q}{Q_0}+\frac{(\Omega_{m,0}-1)}{2-ln(\gamma)}\left(2-ln\left(\gamma \frac{Q}{Q_0}\right)\right)=\frac{\Omega_{m,0}}{a(t)^3}.
\end{equation}
In this case, DE density is expressed as 

\begin{equation}
    2\,\rho_Q=Q_0 \frac{(1-\Omega_{m,0})}{2-ln(\gamma)}\left(2-ln\left(\gamma \frac{Q}{Q_0}\right)\right).
\end{equation}
The DE density parameter is thus found to be
\begin{equation}
    \Omega_Q(t)=\frac{(1-\Omega_{m,0})}{2-ln(\gamma)}\left(2-ln\left(\gamma \frac{Q}{Q_0}\right)\right).
\end{equation}
Using Eq. \eqref{w}, the DE EoS parameter becomes
\begin{equation}
    \omega_Q=-1-\frac{2}{3}\frac{\dot{H}}{H^2}\times \frac{1}{2-ln\left(\gamma \frac{Q}{Q_0}\right)}.
\end{equation}
The deceleration parameter for the logarithmic $F(Q)$ model takes the form
\begin{equation}
    q(t)=-1+\frac{3}{2}\left(\frac{Q+Q_0 \frac{(\Omega_{m,0}-1)}{2-ln(\gamma)}\left(2-ln\left(\gamma \frac{Q}{Q_0}\right)\right)}{Q-\frac{\Omega_{m,0}-1}{2-ln(\gamma)}Q_0}\right).
\end{equation}
For the logarithmic model, the numerical solution of the scale factor is obtained in two ranges ($3.8<\gamma<4.4$ and  $6.5<\gamma<7.3$), which depicts the expansion behavior of the universe. As a result, we choose the integer values of $\gamma$ from these two mentioned intervals as $\gamma=4$ and $\gamma=7$. \\
To obtain the scale factor, we solve Eq. \eqref{38}, assuming $\Omega_{m,0}=0.3$ and $H_0=(14.4)^{-1}$, and we get a numerical solution for $\gamma=4$ and $\gamma=7$. In Fig. \ref{M3}(a), we compare the behavior of the scale factor with the $\Lambda$CDM model, and the model for $\gamma=4$ exhibits good compatibility with the $\Lambda$CDM model as compared to the model for $\gamma=7$. Next, we display the dark energy density parameter $\Omega_Q(t)$ with respect to time $t$ in Fig. \ref{M3}(b), which is not constant. However, the behavior of the DE density parameter is constant in the $\Lambda$CDM model. But, at the present time, the value of $\Omega_Q$ is $0.672$ for $\gamma=4$, which is a good match with the $\Lambda$CDM value, While for $\gamma=7$, the value of $\Omega_Q$ is near about zero, which is not good value. The DE density parameter $\Omega_Q(t)$ eventually converges to $0.9$ for $\gamma=4$ and $1.0$ for $\gamma=7$ at late time.
Also, for both logarithmic models, the total density parameter $\Omega_{Total}$ decreases along with the $\Lambda$CDM model (see in fig. \ref{M3}(c)), and it yields the $\Omega_{Total}(t_0)\approx 1$ at the present time.\\
The evolution trajectory of the DE equation-of-state (EoS) parameter $\omega_{DE}$ with respect to $t$ is shown in Fig. \ref{M3}(d), which is not constant. However, the evolution trajectory of the DE EoS parameter is constant in the $\Lambda$CDM model. 
We can observe in Fig. \ref{M3}(d) that the evolution of the DE EoS parameter began with $-1$ ($\Lambda$CDM model) in the early universe and lies in the non-phantom era. After some time, the DE EoS parameter crosses the phantom divide line and enters the phantom region (in this plot, we are not getting any singularity, it is connected on a large scale of the y-axis). Throughout the early and late time phases, the DE EoS parameter demonstrates $\Lambda$CDM model behavior. The present value of $\omega_{DE}$ is $-1.28$ for $\gamma=4$ and $-1.55$ for $\gamma=7$, and at late-time, it converges to $-1$ ($\Lambda$CDM model).
In Fig. \ref{M3}(e), we can see the evolution trajectory of the deceleration parameter $q(t)$ with respect to time $t$. This trajectory $q(t)$ shows the transition of the universe from the deceleration phase to the acceleration phase. At the present time, it shows the acceleration expansion of the universe, and at late time, it will converge to $-1$.\\
Lastly, we plot the logarithmic model with respect to time $t$ using the obtained scale factor for $\Lambda$CDM case, $\gamma=4$ and $\gamma=7$ (see in Fig. \ref{M3}(f)). Our two logarithmic models decay with time $t$, similar to the $\Lambda$CDM model. As time passes, we observe that the model for $\gamma=4$ exhibits a good match with $\Lambda$CDM compared to the model for $\gamma=7$.

\section{Conclusion}
\label{section 4}
The $f(Q)$ gravity theory has recently garnered much attention as an alternative to dark energy that emanates from the equivalent formulation of GR based on non-metricity: The Symmetric Teleparallel Equivalent of GR (STEGR). In this work, we give theoretical predictions on cosmic observables derived from a modified $f(Q)$ gravity model. We considered the general form of $f(Q)$ model as $f(Q)=Q+F(Q)$, which is equivalent to GR when $F(Q)=0$. To differentiate it from GR, we choose three $F(Q)$ ansatz models, such as power-law, exponential, and logarithmic, and study which model better mimics $\Lambda$CDM evolution from a theoretical point of view. \\
\textit{Power-law model:-} For this model, we try to choose a value of $\lambda$ strictly less than one because it applies to the late-time DE-dominated Universe. For $0<\lambda<1$, the scale factor exhibits expansion behavior, but other cosmological parameters do not demonstrate the good compatibility of the $\Lambda$CDM model. For $\lambda=0$, this power-law model exhibits the $\Lambda$CDM expansion history of the Universe. As a result, we choose two possible non-trivial models for $\lambda=-1$ and $\lambda=-2$. The current values of the DE density parameter $\Omega_Q$ and the total density parameter $\Omega_{Total}$ for both models are approximately 0.697 and 1, respectively. For this model, the DE EoS parameter shows phantom-like behavior, indicating accelerated expansion of the universe. Also, the deceleration parameter shows the universe's present and late-time acceleration expansion. At late-time, both DE EoS and deceleration parameters converge to -1. We also see the behavior of the power-law model versus time $t$. Our two power-law models decay with time $t$, similar to the $\Lambda$CDM model, and we get modest deviations from the $\Lambda$CDM model as time goes on. Finally, we conclude that the negative power of our model shows good compatibility with the $\Lambda$CDM model.\\
\textit{Exponential model:-} Using our current stage of the universe, we found a range for model parameter $\beta$ as $0<\beta<11$ for this model. We studied the behavior of the cosmological parameter for this range of $\beta$. In our study, we choose two possible non-trivial models for $\beta=2$ and $\beta=5$. The current values of the DE density parameter $\Omega_Q$ and the total density parameter $\Omega_{Total}$ for the case of $\beta=2$ are 0.668 and 0.89, respectively, whereas for $\beta=5$ are 0.692 and 0.96, respectively.
For this model, the DE EoS parameter shows phantom-like behavior, indicating accelerated expansion of the universe. Also, the deceleration parameter shows the universe's present and late-time acceleration expansion. At late-time, both DE EoS and deceleration parameters converge to -1. We also see the behavior of the exponential model versus time $t$, which is decay with time $t$, similar to the $\Lambda$CDM model, and we get some modest deviations from $\Lambda$CDM model as time goes on. Finally, we conclude that when we increase the value of $\beta$ from 0 to 11, our exponential model becomes more compatible with the $\Lambda$CDM model.\\

\textit{Logarithmic model:-} Similarly, by using our current stage of the universe, we found two specific ranges of model parameters $\gamma$ as $3.8<\gamma<4.4$ and $6.5<\gamma<7.3$ for this model. In our study, we choose two possible non-trivial models for $\gamma=4$ and $\gamma=7$. The current values of the DE density parameter $\Omega_Q$ and the total density parameter $\Omega_{Total}$ for the case of $\gamma=4$ are approximately 0.672 and 1, respectively, whereas for $\gamma=7$ are near about zero and about one, respectively. For this model, the evolution trajectory of the DE EoS parameter started from the quintessence region and entered the phantom region for both models. 
 Also, the deceleration parameter shows the universe's present and late-time acceleration expansion. At late-time, both DE EoS and deceleration parameters converge to -1. We also seen the behavior of logarithmic model versus time $t$, which is decay with time $t$, similar to the $\Lambda$CDM model, and we observe that the model for $\gamma=4$ exhibits a good match with $\Lambda$CDM compared to the model for $\gamma=7$. As a result, we obtained a range $3.8<\gamma<4.4$ for our logarithmic model, which is might suitable for further study of our universe.

\section*{Data Availability Statement}
There are no new data associated with this article.

\section*{Acknowledgments}
GNG acknowledges University Grants Commission (UGC), New Delhi, India for awarding Junior Research Fellowship (UGC-Ref. No.: 201610122060). PKS acknowledges Science and Engineering Research Board, Department of Science and Technology, Government of India for financial support to carry out Research project No.: CRG/2022/001847 and IUCAA, Pune, India for providing support through the visiting Associateship program. 
We are very much grateful to the honorable referee and to the editor for the illuminating suggestions that have significantly improved our work in terms of research quality, and presentation.


\begin{thebibliography}{90}

\bibitem{Jimenez/2018} J. B. Jimenez et al.,  Phys. Rev. D \textbf{98}, 044048 (2018).
\bibitem{cosmofQ} J. B. Jimenez et al., Phys. Rev. D \textbf{101}, 103507 (2020).

\bibitem{Buchdahl} H. A. Buchdahl, Mon. Not. R. Astron. Soc. \textbf{150}, 1 (1970).

\bibitem{Barrow/1983} J. D. Barrow, A.C. Ottewill, J.Phys. A Math. Gen. \textbf{16}, 2757 (1983).

\bibitem{Capozziello/2002} S. Capozziello, Int. J. Mod. Phys. D \textbf{11}, 483-492 (2002).
\bibitem{Felice/2010} A.  D. Felice and S. Tsujikawa, Living Rev. Rel. \textbf{13}, 3 (2010).



\bibitem{Capozziello/2011} S. Capozziello et al., Phys. Rev. D \textbf{84}, 043527 (2011).

\bibitem{Cai/2016} Yi-Fu Cai et al., Rep. Prog. Phys. \textbf{79}, 106901 (2016).
\bibitem{Farrugia/2016} G. Farrugia and J. Levi Said, Phys. Rev. D \textbf{94}, 124054 (2016).
\bibitem{Paliathanasis/2018} A. Paliathanasis, J. Levi Said, and J. D. Barrow, Phys. Rev. D \textbf{97}, 044008 (2018).
\bibitem{Nester/1999} J. M. Nester, H.-J. Yo, Chin. J. Phys. \textbf{37}, 113 (1999).
\bibitem{Q1} K. F. Dialektopoulos, T. S. Koivisto and S. Capozziello,
Eur. Phys. J. C \textbf{79(7)}, 606 (2019).
\bibitem{Q2} B. J. Barros, T. Barreiro, T. Koivisto and N. J. Nunes,
Phys. Dark Univ. \textbf{30}, 100616 (2020).
\bibitem{Q3} F. Bajardi, D. Vernieri and S. Capozziello, Eur. Phys. J.
Plus \textbf{135}, 912 (2020).
\bibitem{Q4} I. Ayuso, R. Lazkoz and V. Salzano, Phys. Rev. D \textbf{103}, 063505 (2021).
\bibitem{Q5} K. Flathmann and M. Hohmann, Phys. Rev. D \textbf{103},
044030 (2021).
\bibitem{Q6} N. Frusciante, Phys. Rev. D \textbf{103}, 044021 (2021).
\bibitem{Q7} W. Khyllep, A. Paliathanasis and J. Dutta, Phys. Rev. D \textbf{103}, 103521 (2021).
\bibitem{Q8} F. D'Ambrosio, M. Garg and L. Heisenberg, Phys. Lett.
B \textbf{811}, 135970 (2020).
\bibitem{Q9} S. Mandal et al., Phys. Rev. D \textbf{102}, 024057 (2020).

\bibitem{Q11} S. Arora, P. K. Sahoo, Ann. der Phys. \textbf{534}, 2200233 (2022).
\bibitem{Q12} R. Solanki et al., Phys. Dark Univ. \textbf{32}, 100820 (2021).
\bibitem{Q13} S. Mishra, A. Bhat and P. K. Sahoo, EPL \textbf{146}, 29001 (2024).
\bibitem{Q14} D. Rana, R. Solanki, P.K. Sahoo, Phys. Dark Univ. \textbf{43}, 101421 (2024).
\bibitem{ob1} I. Soudi et al., Phys. Rev. D \textbf{100}, 9044008 (2019).
\bibitem{ob2} Ruth Lazkoz et al., Phys. Rev. D \textbf{100}, 104027 (2019).
\bibitem{ob3} Bruno J. Barros et al., Phys. Dark Univ. \textbf{30}, 100616 (2020).
\bibitem{ob4} I. Ayuso, R. Lazkoz, and V. Salzano, Phys. Rev. D \textbf{103}, 063505 (2021).
\bibitem{ob5} F. K. Anagnostopoulos, S. Basilakos, and E. N. Saridakis,  Phys. Lett. B \textbf{822}, 136634 (2021).
\bibitem{ob6} S. Mandal, P. K. Sahoo, Phys. Lett. B \textbf{823}, 136786 (2021).
\bibitem{ob7} L. Atayde, N. Frusciante, Phys. Rev. D \textbf{104}, 064052 (2021).
\bibitem{ob8} N. Frusciante, Phys. Rev. D \textbf{103}, 044021 (2021).
\bibitem{Anagnostopoulos/2023} F. A. Anagnostopoulos et al., Eur. Phys. J. C \textbf{83}, 58 (2023).
\bibitem{simranlss} O. Sokoliuk et al., Mon. Not. Roy. Astron. Soc. \textbf{522}, 252--267 (2023).

\bibitem{gaurav1} G. N. Gadbail et al., Phys. Lett. B \textbf{835}, 137509 (2022).

\bibitem{gaurav2} G. N. Gadbail et al., Eur. Phys. J. C \textbf{83}, 595 (2023).

\bibitem{Esposito/2022} F. Esposito et al., Phys. Rev. D \textbf{105}, 084061 (2022).
\bibitem{gaurav3} G. N. Gadbail, A. De, PK Sahoo, Eur. Phys. J. C \textbf{83}, 1099 (2023).

\bibitem{Beltran/2018} J. B. Jimenez, L. Heisenberg, and T. S. Koivisto, J. Cosmol. Astropart. Phys. \textbf{08}, 039 (2018).

\bibitem{Lazkoz/2019} R. Lazkoz et al., Phys. Rev. D 100, 104027 (2019).


\bibitem{Khyllep/2023} W. Khyllep et al., Phys. Rev. D \textbf{107}, 044022 (2023). 

\bibitem{Ghose/2012} S. Ghose et al., Mon. Not. R. Astron. Soc. \textbf{421}, 20-24 (2012).

\bibitem{Najera/2023} J. A. Najera, C.A. Alvarado, and C. Escamilla-Rivera, Mon. Not. Roy. Astron. Soc. \textbf{524}, 5280--5290 (2023).


\end{thebibliography}
\end{document}